# Forming the COUNCIL Based Clusters in Securing Wireless Ad Hoc Networks

Alok Ojha[1], Hongmei Deng[2], Dharma P. Agrawal[2], and S. Sanyal[3]

[1]Department of Mathematics
IIT, Kharagpur
India 721 302
ojha_iitkgp@yahoo.co.in

[2]Center for Distributed and Mobile Computing
University of Cincinnati, Cincinnati
OH, USA 45221-0030
{hdeng, dpa}@ececs.uc.edu

[3]School of Technology and Computer Sc.
Tata Institute of Fundamental Research
Mumbai, India 400 005
sanyal@tifr.res.in

*Abstract* -- In cluster-based routing protocol (CBRP), two-level hierarchical structure is successfully used to reduce over-flooding in wireless ad hoc networks. As it is vulnerable to a single point of failure, we propose a new adaptive distributed threshold scheme to replace the cluster head by a group of cluster heads within each cluster, called COUNCIL, and distribute the service of single cluster head to multiple cluster heads using $(k,n)$ threshold secret sharing scheme. An ad hoc network formed by COUNCIL based clusters can work correctly when the number of compromised cluster heads is smaller than $k$. To implement this adaptive threshold scheme in wireless ad hoc networks, membership of the clusters should be defined in an adaptive way. In this paper, we mainly discuss our algorithm for forming COUNCIL based clusters using the concept of dominating set from graph theory.

## I. INTRODUCTION

An ad hoc network, by nature, is a highly dynamic network capable of establishing communications without the support of any fixed infrastructure. The underlying self-organizing property makes it useful for many commercial and military applications where geographical or terrestrial constraints demand totally distributed network system. However, the low resource availability and highly dynamically changing topology in these networks necessitates efficient resource utilization and places severe restrictions on designing the routing protocols. Depending on the organization, the routing protocols can be classified as flat or hierarchical [1]. In the flat routing schemes, all the mobile nodes are treated equally, thus the control message need to be propagated in the whole network. If the size of the network is large, then the overhead would be a problem. To mitigate this problem and permit scaling, the hierarchical routing techniques, such as cluster-based routing protocol (CBRP) [2], are applied. But all of these routing protocols works well based on the assumption that all the mobile nodes in the ad hoc networks behaves properly and no compromised, hijacked, or malicious nodes exist in the networks. Obviously this assumption is too strong to be practical.

In this paper, we analyze the vulnerabilities of the CBRP routing protocol, and propose a novel security solution based on the threshold secret sharing scheme. To enhance the fault-tolerance of the network, we replace the cluster head within each cluster by a group of cluster heads, called COUNCIL, and distribute the service of cluster head to multiple cluster heads. The proposed scheme requires adaptively constituting the hierarchical structure in the network. Here, we present a new algorithm for forming the COUNCIL based clusters relying on the availability of the network nodes. The rest of the paper is organized as follows. In Section 2, we give the background and motivation of our work. The proposed algorithm of clustering formation is described in Section 3. The conclusion is added in Section 4.

## II. MOTIVATION

The idea of CBRP comes from the regular cellular networks, which have the support of fixed base stations. In CBRP, the entire network is divided into several overlapped or disjoint clusters. Within each cluster, CBRP replaces the base station by one locally selected cluster head, which has bi-directional links to all its members. Usually the cluster head has more responsibility than the cluster members, such as routing the packets for the cluster members and maintaining the list of cluster members. One natural way for securing CBRP is to give each cluster head more security power such as generating the cluster keys, and keep its private key safely [3]. This type of network structure is vulnerable to problem of single point of failure. If the private key of the cluster head is exposed, then the security scheme becomes meaningless. Or even worse, if the cluster head is compromised, it can choose directing the routing packets to the wrong nodes, collecting the network traffic, modifying the data packets, and even more. That is, it can do everything it could to paralyze the networks. A possible solution to this problem is to distribute the responsibility of cluster head to multiple nodes.

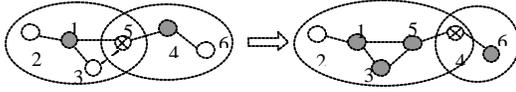

(a) Clusters in CBRP   (b) COUNCIL Based Clusters
Figure 1. Replacing the Cluster Head with COUNCIL

We replace one cluster head by a group of cluster heads with equal functionalities, called COUNCIL, and then deploy the concept of ($k$,$n$) threshold secret sharing [4]. We distribute the cluster head service to multiple cluster heads to enhance the fault-tolerance of the ad hoc networks. That is, the cluster secret is distributed to $n$ COUNCIL nodes, and any $k$ of them can reconstruct the secret, but any collection of less than $k$ partial shares can not get any information about the secret. This allows the network can still work well when the number of compromised cluster heads is smaller than $k$. The requirement for the COUNCIL nodes is that they must be fully connected, i.e., each of them must have bi-directional links to all other nodes in the COUNCIL. Figure 1 demonstrates the proposed idea. Figure 1(a) shows the cluster structure formed by CBRP, in which nodes {1, 4} work as cluster heads, and node 5 connects the two clusters as the gateway node. The dotted line between node 3 and 5 represents that node 3 and 5 are also neighbors of each other, i.e., they can communicate with each other directly. In Figure 1(b), the COUNCIL based clusters are formed. Since the nodes {1, 3, 5} are fully connected, they can work as COUNCIL in the first cluster, and node 6 alone works as the cluster head in the second cluster. The node 4 becomes the gateway node. In the first cluster, the (2, 3) threshold secret sharing scheme can be applied, in which the cluster head service is shared by 3 nodes and any two of them can work together to provide this service.

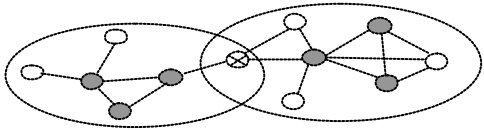

Figure 2. One sample of (2,3) threshold scheme

However, it is difficult to find appropriate value of $n$ since the whole network is not uniformly distributed. If we choose larger value of $n$, then in some local region, it may not be possible to find $n$ cluster heads. While if select smaller value of $n$, smaller size clusters would be formed. Thus, there would be more overhead for the cluster formation, and also in some local area, the network may be denser and we may lose this information by finding only $n$ cluster heads. For example, if we fix the values of $n$ and $k$ to be 3 and 2 respectively. But only one COUNCIL node is selected in the second cluster of Figure 1(b). The (2, 3) threshold secret sharing scheme could not work in this cluster. Figure 2 also demonstrates this issue. Thus, we need to develop a scheme which can find the $n$ value in an adaptive way depending on the availability of the network nodes.

## III. PROPOSED CLUSTERING ALGORITHM

The COUNCIL based clustering algorithm is based on the concept of dominating set in graph theory [5]. Here we first introduce the concept of dominating set, and then present our CABINT based clustering algorithm.

### A. Dominating Set

**Definition:** For a graph $G$ and a subset $D$ of the vertex set $V(G)$, we denote the set of vertices in $G$ which are in $D$ or adjacent to a vertex in $D$ by $N_G[D]$, if $N_G[D] = V(G)$, then $D$ is said to be a dominating set of vertices in $G$.

Now consider the entire ad hoc network as a graph and can be represented by $G = (V, E)$, where $V$ represents a set of wireless mobile nodes and $E$ represents a set of edges between the nodes. An edge between a node pair ($u$,$v$) indicates that node $u$ is within the transmission range of node $v$, and vice visa. Here we assume all the links are bi-directional and the graph is connected. Thus the corresponding graph is an undirected graph. The dominating set of this generated graph is a subset of $V$, called $D$, such that every vertex of $G$ is in $D$ or adjacent to a member of $D$. We require the dominating set be connected, thus it would form the backbone of the network.

**Proposition:** The set of all cluster heads and gateway nodes generated during cluster formation in CBRP forms a dominating set.

**Proof:** Let ? be the set consisting of all the nodes in the entire network. Let $H$ and $G$ denote the sets consisting of all heads and gateway nodes formed during cluster formation in CBRP. Due to the hierarchical structure of CBRP, all the cluster members are only one-hop away from the cluster head, and the gateway node connects adjacent cluster heads. Moreover, all the nodes in CBRP are cluster members, cluster heads, or gateway nodes. Let $D = H \cup G$. Thus, ? is partitioned into nodes that are either within the set $D$, or one-hop away from the members of $S$. Therefore, this set $D$ is a dominating set. Our proposed clustering algorithm is based on the dominating set generated in this way.

## B. Clustering Algorithms

The algorithm runs in two phases. In the first phase, dominating set is generated. Here we use the cluster formation process in CBRP to get the cluster head and gateway nodes and then combine all the cluster heads and the gateway nodes to generate the dominating set. The second phase which immediately follows the first phase, deals with head searching along the dominating set, gateway assignment, and finally forming the clusters.

*(a) Phase1 (Dominating Set Generation)*

Let *V* be the set of mobile nodes, each node has node identity (*NID*), and each cluster formed also has cluster identity (*CID*). *D* stands for the dominating set generated in the first phase. All the nodes broadcast HELLO messages periodically every HELLO_INTERVAL seconds, which contains its neighbor table and cluster adjacent table. The neighbor table contains the *NID* of its neighbor and the role of its neighbor (a cluster head, a cluster member, or a gateway). The cluster adjacent table keeps information about adjacent clusters, which contains the *CID* of neighbor clusters and the gateway node to reach that cluster. Using this broadcast message, each node is able to gather two-hop topology information from itself. The node with lowest *NID* among its neighbor is selected as cluster head, and all its one-hop neighbor nodes become the members of this cluster. Combining the selected cluster heads and gateway, we get the dominating set *D* with size *N*.

*(b) Phase2 (Cluster Formation)*

In phase 2, the algorithm also utilizes the two-hop topology information obtained by each node by exchanging the neighbor table through the HELLO messages. *H* represents the set of cluster heads selected in the entire network and *HC* is the set of cluster heads selected within one cluster. *G* stands for the gateway nodes formed during the second phase. $N(u)$ is the neighbor nodes of node *u*, and $N_d(u)$ stands for the set of neighbor nodes of node *u*, which are also in the dominating set.

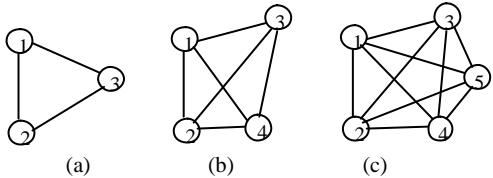

(a)      (b)      (c)
Figure 3. Cliques of size 3, 4, and 5

As mentioned, the cluster heads selected within one cluster must be completely connected to work efficiently since the multiple cluster heads share a same secret and need to exchange information periodically for updating the shares, authenticating the new nodes and encrypting/decrypting messages. They should be completely connected. Selecting the cluster heads is equal to find the completely connected components (clique) within one cluster. It is easy to find a clique of size 3 using the locally 2-hop topology. For an example showed in Figure 3(a), $N(1)=\{2,3\}$, node 1 also collects informtion from its neighbor node 2 that $N(2)=\{1,3\}$, node 1 knows its two-hop neighbor is 3. Since node 3 is

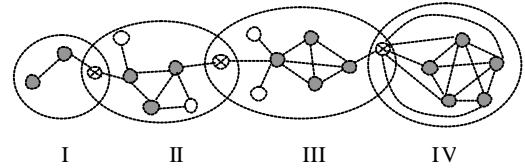

I      II      III      IV

Figure 4. Adaptive cluster formation

also its neighbor, then a triangle is formed from node $\{1, 2, 3\}$. Moreover, any clique with *n* vertices should be shown to be composed of cliques of size 3. In our clustering algorihm, we start from one dominating node, and then use this property to find the completely connected neighbor nodes. So the one dominating node and the completely connected neighbor nodes compose the cluster heads.

Figure 4 shows the clusters formed by the proposed adaptive clustering formalization algorithm given in Figure 5. We have only two cluster heads in cluster I, three cluster heads in cluster II. While we have 4 cluster heads in cluster III and five in cluster IV. Having more cluster heads increases the security level of the ad hoc networks, since the secret of clusters are shared by more nodes. We may have two extreme situations. One situation is that the network is very sparse locally, like cluster 1 in Figure 4, and we can only form the 2-head clusters. In this situation, the minimum security level is obtained since the secret are only shared by two heads. Another situation is that the network is fully connected locally, as cluster IV shows, thus all the fully connected nodes except the gateway node become the cluster heads. We exclude the gateway node since the cluster heads belong to different clusters can not be adjacent. In this case we obtain the maximum security due to the fact that the cluster secret is distributed to all the nodes (except the gateway node) in the cluster. However, this type of cluster formed also becomes unstable, since the shares need to redistribute often by the dynamically changed topology.

## C. Determining the threshold k

After finding the number of cluster heads, which is the value of *n*, we still need to find appropriate threshold *k* ( $1 \leq k \leq n$ ) within each cluster. The threshold *k* is the balance point between the service

```
Function ClusterForm(D)
Input: D(dominating set generated from first phase)
Output: clusters formed
Mark(i=1:N) = FALSE
G:={empty}, H:={empty}
  1) Select h ∈ D, such that h is with lowest NID
  2) HC:={empty}
  3) Mark[h] = TRUE, HC: = HC +{h}, H:=H+{h}
  4) Find a completely connected subset T ⊂ N(h),
     HC: = HC + T, H = H + T. For all the
     nodes s ∈ {T ∩ D}, Mark[s] = TRUE. All the
     nodes (∈ N(H)) form a cluster.
  5) Find g ∈ Nd(s) and g ≠ h, Mark[g]=TRUE,
     G: = G+{g}
  6) D:= D – {(H ∩ D) U G}
  7) Select a new h ∈ Nd(g), and
     Mark(h) = TRUE, repeat the process 2) to 6)
     until Mark(1:N) = TRUE
```

Figure 5. The Adaptive Cluster Formation Algorithm

availability and fault tolerance. We consider two extreme cases.

- *1 out of n*: The secret is shared by $n$ nodes and anyone of them can get the secret. This solution provides maximum availability, but minimal security. As same as the single head solution in CBRP, it is vulnerable to the problem of single point of failure. If one node is compromised, the entire cluster is compromised.
- *n out of n*: The secret is shared by $n$ nodes and only these $n$ nodes working together, the shared secret can be discovered. This solution provides a maximum security, but with minimal availability. If any node happens to be out of service, then the secret can not be discovered.

Usually the practical way is to require $k \geq (n+1)/2$, i.e., a majority of the shareholders is assumed to be honest.

### D. Cluster update and re-formation

In wireless ad hoc networks, each mobile node may move around, thus the network topology changes over time. As a sequence, the dominating set changes, and the clusters formed previously need to update or re-formation. The cluster update means that only small number of clusters update its status locally, while cluster re-formation means that the entire network re-generates the new dominating sets, and also reforms the clusters.

*Cluster update*

Let's consider the situation when a cluster head moves out of the host cluster (the cluster it was in previously) and enters a new cluster, called visiting cluster. The nodes' movement can be viewed as several link disconnections and connections. By periodically exchanging HELLO messages, the link information of its neighbor nodes would be updated. When the moving node enters a new cluster, it may become the cluster members or join cluster head group depending on its relationship with other nodes. If it becomes one of the head, then besides the link information update, the new share is needed to generate for it by the other cluster heads. In this situation, only small changes occur in the network, cluster locally update is a better approach.

*Cluster re-formation*

Whenever more than $n$-$k$ cluster heads move out the host cluster, lot of the gateway nodes moves out of range, the cluster re-formalization is needed. The process starts from the regenerating the dominating set and then form the new clusters. Again, after finishing the cluster formation, the secret shares are also need to redistribute.

## IV. CONCLUSIONS

In this paper, we have proposed an adaptive threshold scheme in securing wireless ad hoc network routing protocol, CBRP. The approach is based on the concept of threshold secret sharing and enhances the fault-tolerance of network by replacing the cluster head with a group of cluster heads. The head's service is distributed to these multiple heads and any $k$ of them can work together to provide the head's service. We have presented an adaptive clustering algorithm to generating the cluster architecture in an adaptive way. Although we make assumptions that all the links between nodes are bi-directional, actually this algorithm also works well in unidirectional links.

ACKNOWLEDGEMENT

This work has been supported by the Ohio Board of Regents Doctoral Enhancement Funds and the National Science Foundation under Grant No. CCR-0113361.